\documentclass[a4paper,11pt]{article}
\usepackage{pos}
\usepackage{mcmule}

\title{Muon-electron scattering at NNLO with \mcmule{}}

\author*[a]{Marco Rocco}

\affiliation[a]{Paul Scherrer Institut \\
CH-5232 Villigen PSI, Switzerland}

\emailAdd{marco.rocco@psi.ch}

\abstract{A recently proposed experiment, MUonE, aims to extract the
  hadronic vacuum polarisation contribution to the muon $g-2$ from
  muon-electron scattering at low energy. The extrapolation requires
  that both experimental and theoretical uncertainties do not exceed
  10 ppm. This corresponds, at least, to
  next-to-next-to-leading-order~(NNLO) QED corrections to $e\mu\to
  e\mu$. I will discuss the implementation of a Monte Carlo integrator
  for this process in the \mcmule{} framework~\cite{Broggio:2022htr},
  which provides infrared-safe differential results at said order in
  QED. An approximation of the MUonE setup provides some
  phenomenological results and sheds light on the need for beyond-NNLO
  corrections, which are currently under study within \mcmule{}.}

\FullConference{%
    16th International Symposium on Radiative Corrections: \\
    Applications of Quantum Field Theory to Phenomenology~(RADCOR2023)\\
    28th May - 2nd June, 2023\\
    Crieff, Scotland, UK
}

\begin{document}
\maketitle

\section{Introduction}
\label{sec:intro}

Excluding new physics, the hadronic vacuum polarisation~(HVP)
contribution to the muon anomalous magnetic moment, $a_\mu = (g-2)/2$,
is generally referred to as the source of the long-standing
discrepancy between experimental measurements~\cite{Muong-2:2006rrc,
  Muong-2:2021ojo, Muong-2:2023cdq} and Standard Model~(SM)
predictions. Currently, there is not an agreement among the latter, as
predictions using lattice QCD differ from those employing data-driven
dispersive calculations. In contrast to dispersive
predictions~\cite{Aoyama:2020ynm}, whose discrepancy with the
experiment amounts up to 5$\sigma$, a recent calculation in lattice
QCD~\cite{Borsanyi:2020mff} drastically reduces the discrepancy. In
this scenario, a different approach to a data-driven calculation,
expressed e.g.~by the MUonE experiment~\cite{Abbiendi:2016xup,
  CarloniCalame:2015obs}, is decisive to further disentangle the
problem.

Traditional dispersive calculations rely on experimental inputs from
$e^+ e^- \to {\rm hadrons}$, measured in the time-like region ($s>0$)
at energies around 1-10 GeV, where numerous hadronic resonances hamper
the experimental precision.

\begin{figure}[ht!]
    \centering
    \includegraphics[width=0.9\textwidth]{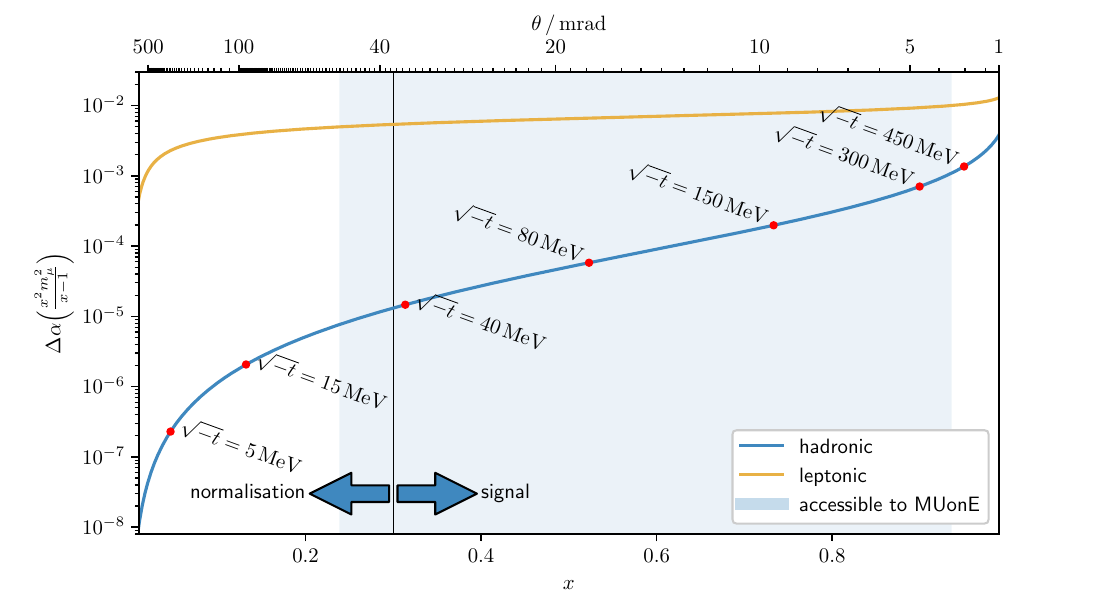}
    \caption{HVP contribution to the running of $\alpha$ for
      space-like kinematics, computed as the NLO correction to
      muon-electron scattering due to HVP insertions, as a function of
      $x$ and $\theta_e$. The contribution due to leptonic VP
      insertions is also given for comparison, along with the
      kinematic region accessible to MUonE.}
    \label{fig:alpha-x}
\end{figure}

On the other hand, the MUonE experiment consists of a 160 GeV muon
beam colliding on a fixed target of atomic electrons, in a pure
$t$-channel. From the measurement of the scattering angles of muons
and electrons, $\theta_\mu$ and $\theta_e$, in elastic events, the HVP
contribution to the running of the electromagnetic coupling,
$\alpha(t(x)<0)$, can be reconstructed via a template fit in the
space-like region, where no resonance hampers the measurement. The
formula~\cite{CarloniCalame:2015obs}
\begin{equation}
  a_\mu^{\rm HVP} = \frac{\alpha}{\pi} \int_0^1 \D x
  (1-x)\,\Delta\alpha^{\rm had}(t(x)) \,,\qquad t(x) = -\frac{x^2 \,
    m_\mu^2}{1-x} \,,
\end{equation}
yields the HVP contribution to the muon
anomaly. Figure~\ref{fig:alpha-x} shows a simulation of the running of
$\alpha$, split into a leptonic and a hadronic part. The former is
computed perturbatively, while the second employs the library {\tt
  alphaQED}~\cite{alphaqed}.

The kinematics of MUonE is particularly favourable, as the area
accessible to the experiment covers most of the HVP contribution in
Figure~\ref{fig:alpha-x}, i.e.~corresponding to higher values of $x$,
or equivalently smaller values of $\theta_e$. Further, this allows to
define a normalisation region, where the HVP signal is much lower.

However, the accuracy of the total experimental and theoretical error
should not exceed 10 ppm, as the signal of the experiment is ${\cal
  O}(10^{-3})$, and the HVP needs to be extracted with a precision
below one percent, in order to match the statistical error of the
other evaluations.

On the theoretical side, there has been a coordinated
effort~\cite{Banerjee:2020tdt} aiming at developing two completely
independent Monte Carlo event generators for muon-electron
scattering. At least an NNLO calculation in QED is mandatory to reach
the precision required, and the results at this order suggest the need
for the resummation of large logarithms and the calculation of~(the
dominant part of) the N$^3$LO corrections. In addition to higher-order
QED effects, nuclear effects as well as pion and lepton-pair
production have to be taken into account~\cite{Budassi:2022kqs}, but
will not be considered in this contribution.

The {\sc Mesmer} Monte Carlo provides the complete set of electroweak
NLO corrections~\cite{Alacevich:2018vez}, as well as QED NNLO
corrections~\cite{CarloniCalame:2020yoz, Budassi:2021twh}, using an
approximation for genuine two-loop four-point topologies. In order to
cope with infrared divergences photon-mass regularisation combined
with a slicing approach is employed. In parallel, the \mcmule{}
framework~\cite{mcmule} offers another efficient environment to reach
the desired level of accuracy. Section~\ref{sec:mcmule} discusses the
implementation in \mcmule{} of muon-electron scattering at NNLO in
QED, which can serve as a Monte Carlo generator~(at present,
integrator) for the MUonE experiment. A subset of the results is
presented in Section~\ref{sec:results}, before concluding in
Section~\ref{sec:conclusions}.

\section{\mcmule{} for MUonE}
\label{sec:mcmule}

The implementation of muon-electron scattering at NNLO in the
\mcmule{} framework, along with results and discussion, is presented
in greater detail in~\cite{Broggio:2022htr}
and~\cite{Engel:2022kde}. Here follows a summary, with a particular
focus on phenomenology.

The general idea behind \mcmule{} is the adaptation of techniques
developed and employed in the context of higher-order perturbative
QCD, to higher-order studies in QED. For example, \mcmule{} uses
dimensional regularisation and a generalisation of the FKS
method~\cite{Frixione:1995ms, Frederix:2009yq} to any order in the
electromagnetic coupling, FKS$^\ell$~\cite{Engel:2019nfw}, for the
subtraction of soft divergences. In fact, MUonE observables are not
collinear safe, and therefore strongly depend on fermion-mass
effects. Thus, for fully differential predictions, taking those into
account is strictly necessary. At the same time, lepton masses
regulate collinear divergences, leaving soft divergences only.

Further, the same machinery developed to handle one- and multi-loop
integrals can be adapted to QED, where the presence of additional
scales, such as the lepton masses, makes loop integrations more
difficult. For one-loop problems, \mcmule{} employs {\sc
  OpenLoops}~\cite{Buccioni:2017yxi, Buccioni:2019sur}, which proved
to be remarkably stable except for phase-space regions where a photon
is particularly soft, or a pseudo-collinear configuration leads to the
presence of large logarithms. A good numerical stability is recovered
using next-to-soft~(NTS) stabilisation~\cite{Banerjee:2021mty,
  Engel:2021ccn} for such regions, i.e.~employing the leading- and
next-to-leading-soft terms in the photon-energy expansion of the
relevant matrix element, instead of the full matrix element. With
two-loop diagrams no automatic procedure is able to deal with generic
processes. It is then necessary to resort to external results that
consider the particular process of interest. These results are often
built for QCD, where the mass of the light quarks can be
neglected. Thus, \mcmule{} employs calculations where the mass of the
fermion is neglected, but can recover those neglected effects via
massification~\cite{Penin:2005eh, Becher:2007cu, Engel:2018fsb}.

A more detailed discussion about the methods used in \mcmule{}~(and a
validation of them) can be found in the original muon-electron
scattering paper~\cite{Broggio:2022htr}, or elsewhere in these
proceedings~\cite{Ulrich:2023mule}. For the purpose of the present
contribution, it is sufficient to say that muon-electron scattering at
NNLO is implemented as follows. Contributions at NLO~(real and
virtual) and NNLO~(double-real, real-virtual and double-virtual) are
divided into photonic and fermionic. The latter are those containing a
fermionic vacuum polarisation insertion, and were calculated using the
hyperspherical method~\cite{Fael:2018dmz}, then validated by a second
calculation done via a dispersive method~\cite{Fael:2019nsf}.

In general, both photonic and fermionic contributions can be
subdivided into three gauge-invariant subsets according to their
formal leptonic charge. As for the sample diagrams below,
contributions where photon radiation only attaches to the
electron~(muon) line are called electronic~(muonic), all other
contributions are referred to as mixed. In the \mcmule{} framework,
each contribution can be computed separately, allowing the user to
study the impact of different classes and to infer possible
hierarchies among them.
\begin{eqnarray}
  & {\rm electronic} & \qquad
  \begin{tikzpicture}[scale=.3,baseline={(0,0.2)}]
        
    \draw[line width=.2mm]  (-1.5,0) -- (1.5,0);
    \draw[line width=.2mm]  (-1.5,-.1) -- (1.5,-.1);
    \draw [line width=0.2mm, tightphoton] (0,2.5)--(0,0);
    \draw[line width=.2mm]  (-1.5,2.5) -- (1.5,2.5);
    \centerarc [line width=0.2mm,tightphoton](-0.3,2.5)(0:180:.9)
    \centerarc [line width=0.2mm,tightphoton](0.3,2.5)(0:180:.9)

  \end{tikzpicture}
  \hspace{.5cm}
  \begin{tikzpicture}[scale=.3,baseline={(0,0.2)}]
        
    \draw[line width=.2mm]  (-1.5,0) -- (1.5,0);
    \draw[line width=.2mm]  (-1.5,-.1) -- (1.5,-.1);
    \draw [tightphoton, line width=0.2mm] (0,2.5)--(0,0);
    \draw [tightphoton, line width=0.2mm] (0.5,2.5)--(1.5,3.5);
    \draw[line width=.2mm]  (-1.5,2.5) -- (1.5,2.5);
    \centerarc [line width=0.2mm,tightphoton](0,2.5)(0:180:0.9)

  \end{tikzpicture}
  \hspace{.5cm}
  \begin{tikzpicture}[scale=.3,baseline={(0,0.2)}]
        
    \draw[line width=.2mm]  (-1.5,0) -- (1.5,0);
    \draw[line width=.2mm]  (-1.5,-.1) -- (1.5,-.1);
    \draw [tightphoton, line width=0.2mm] (0,2.5)--(0,0);
    \draw [tightphoton, line width=0.2mm] (-1,2.5)--(0,3.5);
    \draw [tightphoton, line width=0.2mm] (.5,2.5)--(1.5,3.5);
    \draw[line width=.2mm]  (-1.5,2.5) -- (1.5,2.5);

    \draw [tightphoton, line width=0.2mm] (.5,2.5)--(1.5,3.5);

  \end{tikzpicture} \nonumber\\[2mm]
  & {\rm mixed} & \qquad
  \begin{tikzpicture}[scale=.3,baseline={(0,0.2)}]
        
    \draw[line width=.2mm]  (-1.5,0) -- (1.5,0);
    \draw[line width=.2mm]  (-1.5,-.1) -- (1.5,-.1);
    \draw [line width=0.2mm, tightphoton] (-1,2.5)--(0,0);
    \draw [line width=0.2mm, tightphoton] (0,2.5)--(-1,0);
    \draw [line width=0.2mm, tightphoton] (1,2.5)--(1,0);
    \draw[line width=.2mm]  (-1.5,2.5) -- (1.5,2.5);

  \end{tikzpicture}
  \hspace{.5cm}
  \begin{tikzpicture}[scale=.3,baseline={(0,0.2)}]
        
    \draw[line width=.2mm]  (-1.5,0) -- (1.5,0);
    \draw[line width=.2mm]  (-1.5,-.1) -- (1.5,-.1);
    \draw [tightphoton, line width=0.2mm] (-.8,2.5)--(-.8,0);
    \draw [tightphoton, line width=0.2mm] ( .8,2.5)--( .8,0);
    \draw [tightphoton, line width=0.2mm] (0.5,2.5)--(1.5,3.5);
    \draw[line width=.2mm]  (-1.5,2.5) -- (1.5,2.5);

  \end{tikzpicture}
  \hspace{.5cm}
  \begin{tikzpicture}[scale=.3,baseline={(0,0.2)}]
        
    \draw[line width=.2mm]  (-1.5,0) -- (1.5,0);
    \draw[line width=.2mm]  (-1.5,-.1) -- (1.5,-.1);
    \draw [tightphoton, line width=0.2mm] (0,2.5)--(0,0);
    \draw [tightphoton, line width=0.2mm] (-1,2.5)--(0,3.5);
    \draw [tightphoton, line width=0.2mm] (.5,0)--(1.5,-1.);
    \draw[line width=.2mm]  (-1.5,2.5) -- (1.5,2.5);


  \end{tikzpicture} \nonumber\\[2mm]
  & {\rm muonic} & \qquad
  \begin{tikzpicture}[scale=.3,baseline={(0,0.2)}]
        
    \draw[line width=.2mm]  (-1.5,0) -- (1.5,0);
    \draw[line width=.2mm]  (-1.5,-.1) -- (1.5,-.1);
    \draw [line width=0.2mm, tightphoton] (0,2.5)--(0,0);
    \draw[line width=.2mm]  (-1.5,2.5) -- (1.5,2.5);
    \centerarc [line width=0.2mm,tightphoton](-0.3,0)(180:360:.9)
    \centerarc [line width=0.2mm,tightphoton](0.3,0)(180:360:.9)

  \end{tikzpicture}
  \hspace{.5cm}
  \begin{tikzpicture}[scale=.3,baseline={(0,0.2)}]
        
    \draw[line width=.2mm]  (-1.5,0) -- (1.5,0);
    \draw[line width=.2mm]  (-1.5,-.1) -- (1.5,-.1);
    \draw [tightphoton, line width=0.2mm] (0,2.5)--(0,0);
    \draw [tightphoton, line width=0.2mm] (0.5,0)--(1.5,-1.5);
    \draw[line width=.2mm]  (-1.5,2.5) -- (1.5,2.5);
    \centerarc [line width=0.2mm,tightphoton](0,0)(180:360:0.9)

  \end{tikzpicture}
  \hspace{.5cm}
  \begin{tikzpicture}[scale=.3,baseline={(0,0.2)}]
        
    \draw[line width=.2mm]  (-1.5,0) -- (1.5,0);
    \draw[line width=.2mm]  (-1.5,-.1) -- (1.5,-.1);
    \draw [tightphoton, line width=0.2mm] (0,2.5)--(0,0);
    \draw [tightphoton, line width=0.2mm] (-1,0)--(0,-1.);
    \draw [tightphoton, line width=0.2mm] (.5,0)--(1.5,-1.);
    \draw[line width=.2mm]  (-1.5,2.5) -- (1.5,2.5);


  \end{tikzpicture} \nonumber
\end{eqnarray}

Double-virtual electronic and muonic contributions were computed with
full mass dependence, using the analytic expressions for the heavy
quark form factors of~\cite{Bernreuther:2004ih}, while mixed
double-virtual contributions, which involve genuine two-loop
four-point topologies, were computed applying massification to the
results of~\cite{Bonciani:2021okt, Mandal:2022vju}, which employ the
master integrals computed in~\cite{Mastrolia:2017pfy, DiVita:2018nnh,
  Mandal:2022vju}. This is the only approximation made in the
\mcmule{} prediction, amounting to the neglect, at NNLO, of terms that
are polynomially suppressed in the electron-mass expansion of the
double-virtual contribution. NTS stabilisation was then applied to all
real-virtual contributions, in order to achieve the desired numerical
stability. As shown in the original paper, the use of NTS expansions
results in the neglect of terms that are much below the 10 ppm
requirement by MUonE.

\begin{figure}[ht!]
    \centering
    \includegraphics[width=0.9\textwidth]{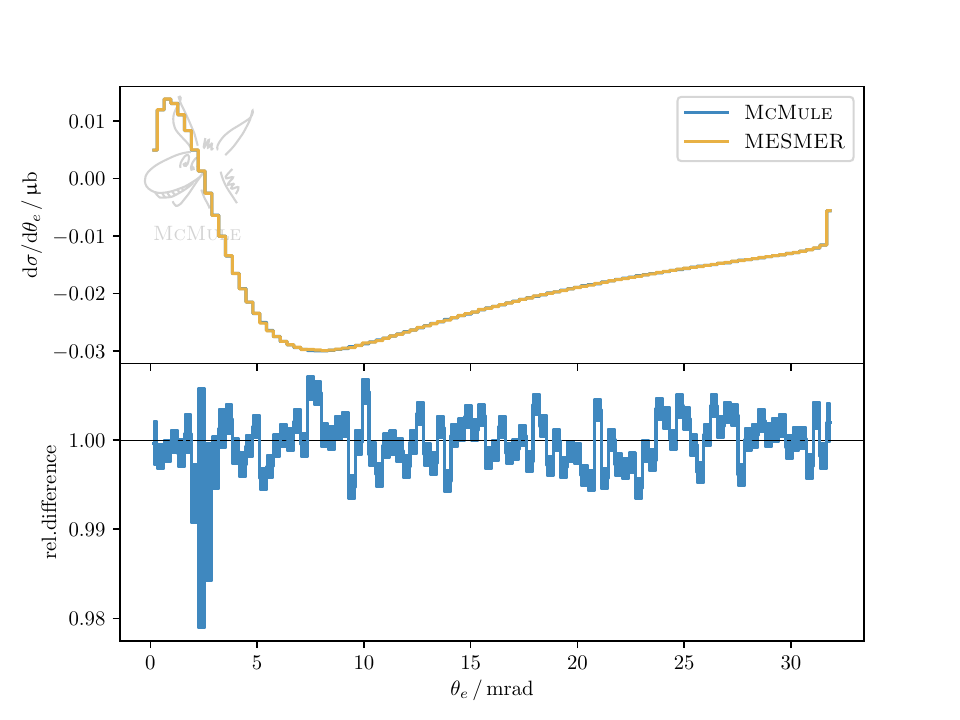}
    \caption{Top: NNLO double-real and real-virtual contributions to
      muon-electron scattering as differential distributions
      w.r.t.~$\theta_e$, computed by {\sc Mesmer}~(yellow) and
      \mcmule{}~(blue). A small photon energy cut of $10^{-6} \times
      \sqrt{s}/2$ was used. Bottom: Ratio between the two
      predictions. The larger oscillation is unphysical, corresponding
      to the zero crossing of the distributions.}
    \label{fig:check}
\end{figure}

A number of internal and external tests were carried out in order to
validate the results, cf.~Section 4 of~\cite{Broggio:2022htr}. Here we
comment on the comparison of the mixed contributions to the photonic
NNLO correction, which have been calculated both in {\sc Mesmer} and
in \mcmule{}. Since the two frameworks employ different scheme to
handle IR divergences, a comparison between the two results represents
a completely independent check. As the calculation
in~\cite{Budassi:2021twh} is complete up to the mixed two-loop
contribution, it is possible to compare the mixed NLO correction to
$\mu e \to \mu e \gamma$, which is physical and corresponds to the
double-real and real-virtual contributions to muon-electron
scattering. In order to check the numerical stability of the
real-virtual implementation, small photon energy cuts of
$\{10^{-6},\,10^{-5},\,10^{-4}\} \times \sqrt{s}/2$ were used. Perfect
agreement was found between the two codes for the total cross section
as well as for differential distributions, as shown in
Figure~\ref{fig:check}, at sub-percent level.

\section{Results}
\label{sec:results}

This section presents some results for muon-electron scattering at
NNLO, with the characteristics of the MUonE experiment in mind.  The
kinematics is defined by
\begin{align}
\label{eq:kinematics}
  e^-(p_1)\, \mu^\pm(p_2) \to e^-(p_3, E_e, \theta_e)\, \mu^\pm(p_4,
  E_\mu, \theta_\mu) + \{\gamma(k_1)\, \gamma(k_2)\} \, ,
\end{align}
with the outgoing electron and muon energy, $E_e$ and $E_\mu$, and the
electron and muon scattering angle with respect to the beam axis,
$\theta_e$ and $\theta_\mu$. The muon beam energy is set to 160 GeV,
consistent with the M2 beam line at CERN North
Area~\cite{Spedicato:2022qtw}. A cut is imposed on the energy of the
outgoing electron, $E_e > 1$ GeV, which is equivalent to a cut on the
minimal value of $|t|$, in order to cure the singular behaviour of
$\D\sigma/\D t \sim t^{-2}$, where $t$ is the usual Mandelstam
invariant. A cut on $\theta_\mu$ can be used to remove most of the
background. Hence, for the results shown here, $\theta_\mu > 0.3$ mrad
was also required.

Given this kinematical setup, \mcmule{} is able to produce
differential distributions for any infrared-safe observable. At
present, it does not generate events and can only act as a Monte Carlo
integrator. However, the possibility to generate events will be
available in the near future~\cite{Ulrich:evgen}.

In this contribution, the focus is on differential distributions
w.r.t.~the scattering angle of the electron, as this is the main
interest of the MUonE experiment, in particular for a beam of negative
muons. The whole set of results is instead presented in the original
paper and publicly available at the \mcmule{} Zenodo
repository~\cite{McMule:data}.

\begin{figure}[ht!]
    \centering
    \includegraphics[width=0.9\textwidth]{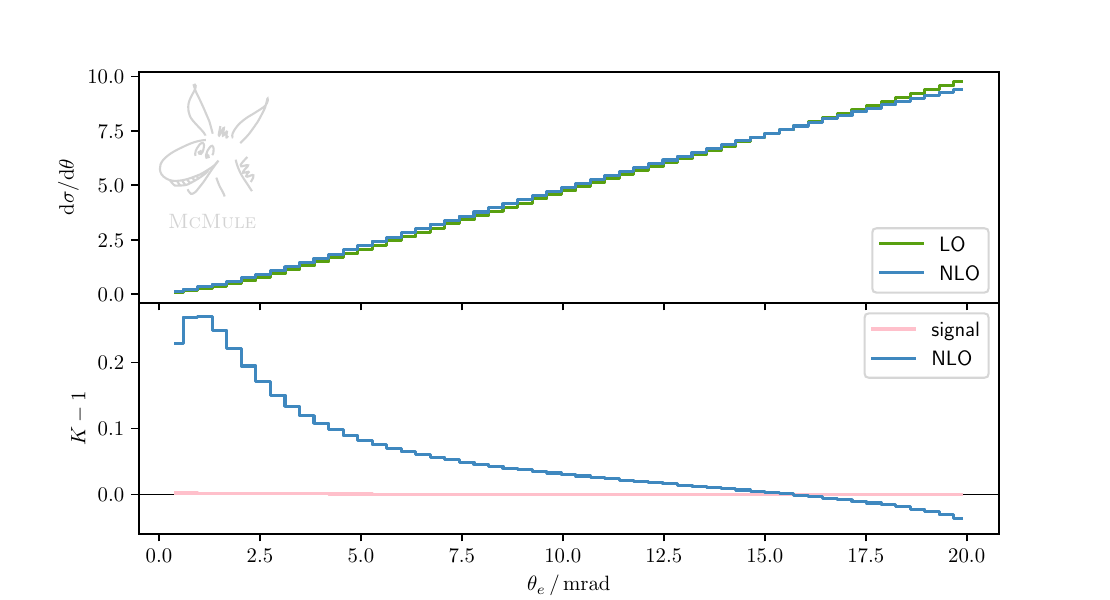}
    \caption{Top: LO~(green) and NLO~(blue) differential cross section
      w.r.t.~$\theta_e$. Bottom: $K$ factors of the NLO correction and
      the signal of the MUonE experiment~(pink), i.e.~the NLO HVP
      contribution.}
    \label{fig:the-s1-nlo}
\end{figure}

\begin{figure}[ht!]
    \centering
    \includegraphics[width=0.9\textwidth]{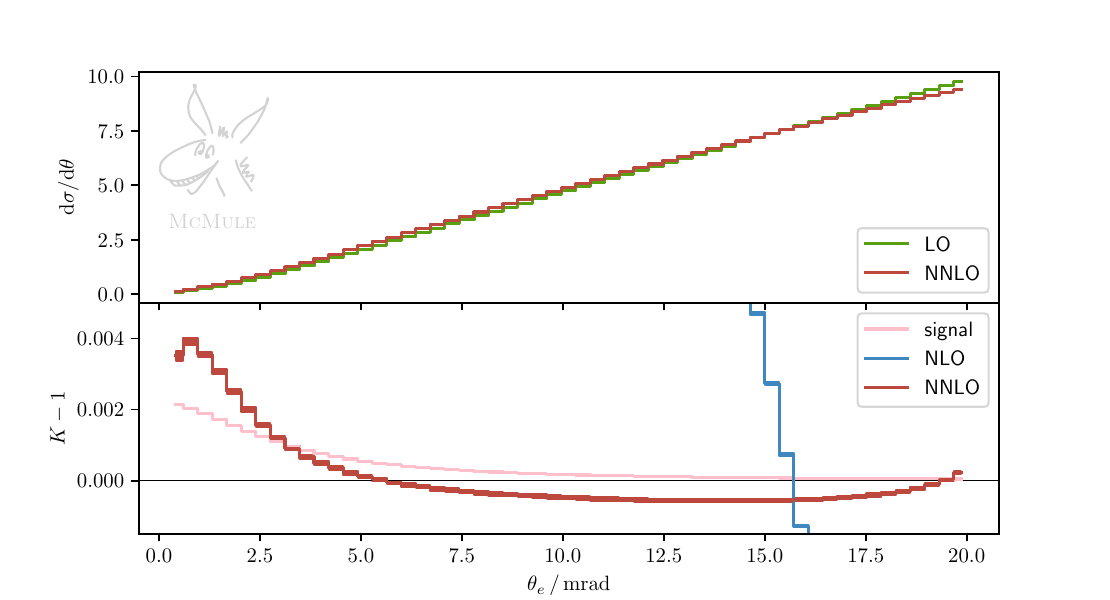}
    \caption{Top: LO~(green) and NNLO~(red) differential cross section
      w.r.t.~$\theta_e$. Bottom: $K$ factors of the NLO and NNLO
      correction, and the signal of the MUonE experiment~(pink),
      i.e.~the NLO HVP contribution.}
    \label{fig:the-s1-nnlo}
\end{figure}

Figure~\ref{fig:the-s1-nlo} and~\ref{fig:the-s1-nnlo} show, in the
upper panel, the LO and (N)NLO angular distributions, and, in the
lower panel, the $K$ factor for the NLO and NNLO distributions,
defined as
\begin{align}
    K^{(i)} - 1 = \frac{\sigma_i}{\sigma_{i-1}} \,,
\end{align}
where $\sigma_k = \sum_{i=0}^k \sigma^{(i)}$ is given by the sum of
the order-by-order contributions, $\sigma^{(i)}$, to the N$^{k}$LO
integrated cross section. In addition, the $K$ factor of the signal is
also shown, corresponding to the hadronic part of the NLO fermionic
contribution.

(N)NLO corrections amount up to 20\% (0.2\%), particularly for small
electron scattering angles, or equivalently for large electron
energies, where photon emission is forced to be soft. In this
kinematical configuration, the signal is completely outweighed by the
NLO correction, and turns out to be of the same order or smaller than
the NNLO correction. Further, the enhancement observed in the
small-$\theta_e$ region suggests the need for a more reliable
description of the region where large logarithms cause such behaviour.

\begin{figure}[ht!]
    \centering
    \includegraphics[width=0.9\textwidth]{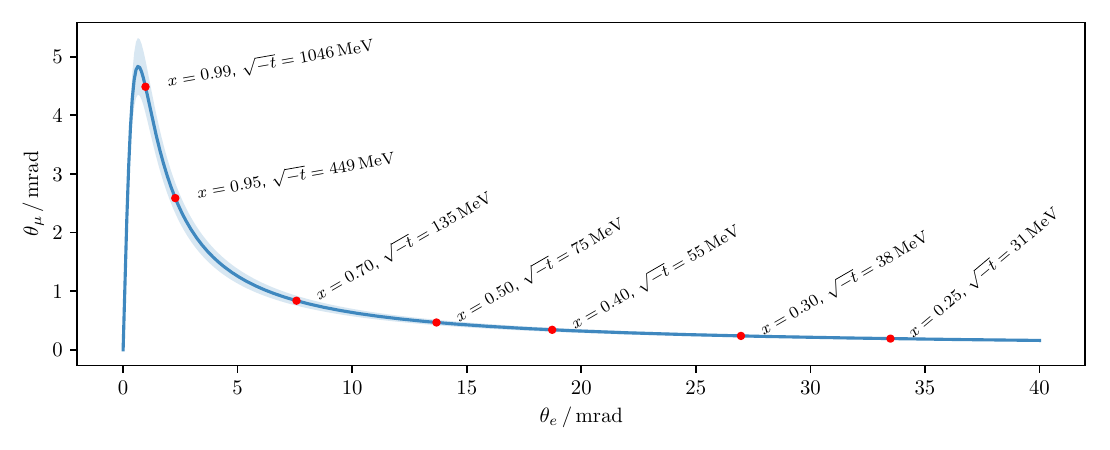}
    \caption{$\theta_\mu$ as a function of $\theta_e$ for elastic
      events. The light-blue band corresponds to the elasticity cut.}
    \label{fig:ela}
\end{figure}

\begin{figure}[ht!]
    \centering
    \includegraphics[width=0.9\textwidth]{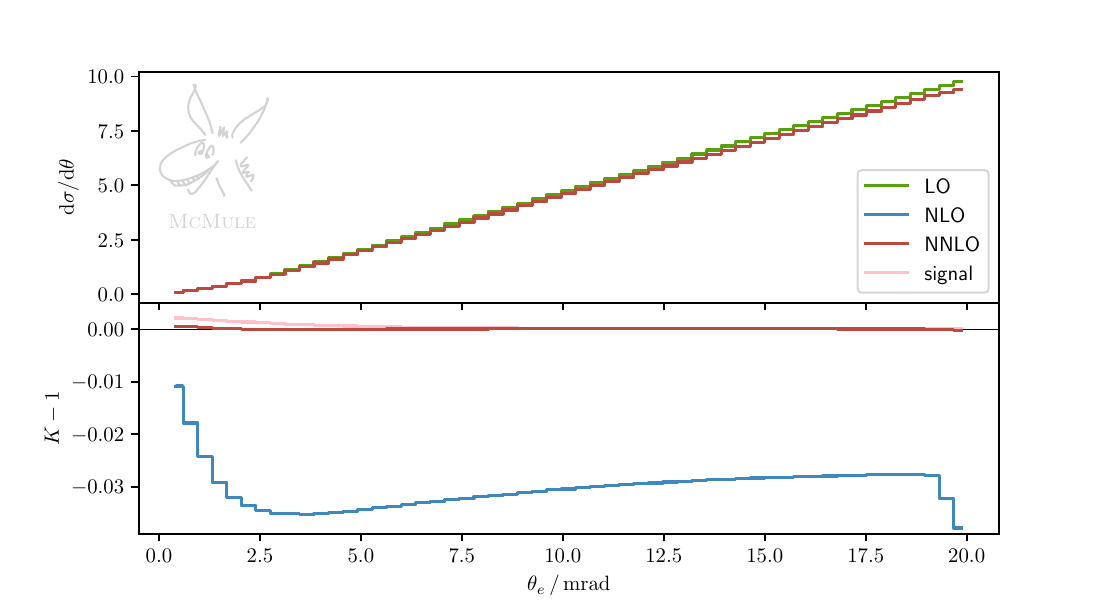}
    \caption{Top: LO~(green) and NNLO~(red) differential cross section
      w.r.t.~$\theta_e$. Bottom: $K$ factors of the NLO and NNLO
      correction, and the signal of the MUonE experiment~(pink),
      i.e.~the NLO HVP contribution. All curves are obtained after
      applying the elasticity cut discussed in the text.}
    \label{fig:the-s2}
\end{figure}

In order to achieve a well-defined extraction of the signal, not
hampered by such dominant QED corrections in the background, a
possible way to proceed is to discriminate elastic scattering events
from the otherwise kinematically allowed radiative events and
processes. This can be obtained in terms of the elasticity constraint
that relates muon and electron scattering angles in the absence of
photons, shown in Figure~\ref{fig:ela}.
For example, the requirement $0.9 < \theta_\mu/\theta^{\rm el}_\mu <
1.1$, where $\theta^{\rm el}_\mu$ is the muon scattering angle
as defined in Figure~\ref{fig:ela} as a function of the electron
scattering angle, can act as a veto for hard radiation. The angular
distributions in the presence of this additional elasticity cut are
displayed in Figure~\ref{fig:the-s2}.

As expected from an evenly-distributed soft enhancement, the $K$
factor is significantly reduced and flattened. In this context, the
NLO correction can be subtracted more efficiently, and the signal can
be extracted on top of the NNLO correction, which is now, in general,
smaller. However, such a kinematical constraint is not ideal from the
experimental perspective. It would cut off many events, yielding
issues in terms of statistics, and would also complicate the estimate
of systematic uncertainties, as it would lead to a complex practical
implementation. At present, the alternative proposed by the experiment
is to employ a template fit to extract the HVP, as discussed
in~\cite{Abbiendi:2022oks}. Nonetheless, a study with the elasticity
cut is still of theoretical interest.

\section{Conclusions and outlook}
\label{sec:conclusions}

We have reviewed the implementation in the \mcmule{} framework of
muon-electron scattering at NNLO in QED, as well as some of the
results presented therein. This corresponds to the first result at
NNLO for a two-to-two process in QED with two different non-vanishing
masses on the external lines. A more detailed description of the
methods employed can be found elsewhere in these
proceedings~\cite{Ulrich:2023mule}.

The MUonE experiment may benefit from these results for the extraction
of the HVP contribution to the muon $g-2$. The \mcmule{} effort is
part of a bigger theoretical effort~\cite{Banerjee:2020tdt}, whose aim
is to provide the most precise prediction for muon-electron
scattering, in order to match the 10 ppm precision goal. The magnitude
of the NNLO corrections at differential level, around $10^{-3}$, is
still too large compared to said goal. Thus, higher-order predictions
beyond NNLO can certainly help in that direction. Furthermore, it is
mandatory to make an effort towards a more reliable
description of the region where radiation leads to an enhancement
through large logarithms.

An N$^3$LO prediction~(or at least the dominant part of it) and a more
precise description of large-log regions, through resummation or via
the implementation of a parton shower, are on the agenda of an ongoing
effort, started with a series of workstops in 2022 and
2023~\cite{Durham:n3lo, Zuerich:n3lo}.

\paragraph{Acknowledgement}
I acknowledge support from the Swiss National Science
Foundation~(SNSF) under grant 200020\_20738. A huge thank you to all
the colleagues of the original paper~\cite{Broggio:2022htr}, upon
which this contribution is mainly based.

\bibliographystyle{JHEP}
\bibliography{biblist}

\end{document}